\documentclass[12pt]{article}

\usepackage{latexsym}
\usepackage{amssymb,amsfonts,amsmath}
\usepackage{graphicx} 
\usepackage{indentfirst}
\usepackage{bbm}
\usepackage{amssymb}
\usepackage{verbatim}
\usepackage{amsmath, amsthm,amssymb}
\usepackage{mathrsfs}
\usepackage{hyperref}
\usepackage{amsfonts}
\usepackage{dsfont}
\usepackage{cite}
\usepackage{xcolor}
\usepackage[multiple]{footmisc}
\parskip=\medskipamount

\newcommand {\cD}{{\cal D}}

\newcommand {\cN}{{\cal N}}

\def\a{\alpha}

\def\b{\beta}

\def\d{\delta}
\def\e{\epsilon}

\def\j{\psi}

\def\l{\lambda}

\def\o{\omega}

\def\s{\sigma}

\def\J{\Psi}
\def\L{\Lambda}
\def\O{\Omega}

\def\Q{\Theta}
\def\S{\Sigma}

\def\tr{{\rm tr}}
\def\rd{{\rm d}}
\def\ri{{\rm i}}

\newcommand{\ad}{{\dot{\alpha}}}                           %new
\newcommand{\bd}{{\dot{\beta}}}                            %new
\newcommand{\ve}{\varepsilon}                            %new
\newcommand{\hf}{\frac12}
\newcommand{\be}{\begin{equation}}
\newcommand{\ee}{\end{equation}}
\newcommand{\bea}{\begin{eqnarray}}
\newcommand{\eea}{\end{eqnarray}}
\newcommand{\non}{\nonumber}
\newcommand{\bm}[1]{\mbox{\boldmath$#1$}}

\def\double #1{#1{\hbox{\kern-2pt $#1$}}}

\newif\ifdtup

\newcommand{\bsubeq}{\begin{subequations}}
\newcommand{\esubeq}{\end{subequations}}
\numberwithin{equation}{section}

\newcommand{\sSU}{\mathsf{SU}}
\newcommand{\sSL}{\mathsf{SL}}

\newcommand{\sO}{\mathsf{O}}
\newcommand{\sU}{\mathsf{U}}

\newcommand{\sISO}{\mathsf{ISO}}
\begin{document}

\begin{titlepage}
%\begin{flushright}
%October, 2021 \\
%Revised version: May, 2022
%\end{flushright}
%\vspace{5mm}

\begin{center}
{\Large \bf 
Local supersymmetry: Variations on a theme by Volkov and Soroka} 
\end{center}

\begin{center}

{\bf Sergei M. Kuzenko} \\
\vspace{5mm}

\footnotesize{
{\it Department of Physics M013, The University of Western Australia\\
35 Stirling Highway, Perth W.A. 6009, Australia}}  
~\\
\vspace{2mm}
~\\
Email: \texttt{ 
sergei.kuzenko@uwa.edu.au}\\
\vspace{2mm}

\end{center}

\begin{abstract}
\baselineskip=14pt
We revisit the work by Volkov and Soroka on spontaneously broken local supersymmetry. It is demonstrated for the first time that, for specially chosen parameters of the theory, the Volkov-Soroka action is invariant under two different local supersymmetries. One of them is present for arbitrary values of the parameters and acts on the Goldstino, while the other supersymmetry emerges only in a special case and leaves the Goldstino invariant. The former can be used to gauge away the Goldstino, and then the resulting action coincides with that proposed by Deser and Zumino for consistent supergravity in the first-order formalism. In this  sense, pure $\mathcal{N} = 1$ supergravity is a special case of the Volkov-Soroka theory, although it was not discovered by these authors.
 We also explain how the Volkov-Soroka approach  allows one to naturally arrive at the 1.5 formalism.
 Our analysis provides a nonlinear realisation approach to construct unbroken $\mathcal{N} = 1$ Poincar\'e supergravity. 
\end{abstract}
\vspace{5mm}

\vfill

\vfill
\end{titlepage}

\newpage
\renewcommand{\thefootnote}{\arabic{footnote}}
\setcounter{footnote}{0}

%
%\tableofcontents{}
%\vspace{1cm}
%\bigskip\hrule
%

\allowdisplaybreaks

%%%%%%%%%%%%%%%%%%%%%%%%%%%%%%%%
%%%%%%%%%%%%%%%%%%%%%%%%%%%%%%%%

\section{Introduction}

Supergravity was discovered in 1976 \cite{FvNF,DZ}. In 1994, less than two years before his death, Volkov posted two preprints to the hep-th archive \cite{Volkov1,Volkov2}, both of which contained ``supergravity before 1976'' in the title. 
He argued that the crucial ingredients of $\cN=1$ supergravity in four dimensions had appeared in his earlier work with Soroka \cite{VS,VS2}.\footnote{Similar thoughts were also expressed by Soroka a  few years later \cite{Soroka1,Soroka2}, see also \cite{Nurmagambetov, Dup}.} Since the concept of local supersymmetry has played a fundamental role in modern theoretical physics, it is suitable to have a fresh critical look at the Volkov-Soroka construction.

In this paper we revisit the Volkov-Soroka approach to spontaneously broken local supersymmetry  \cite{VS,VS2}.
We demonstrate for the first time that, for specially chosen parameters of the theory, the Volkov-Soroka action is invariant under two different local supersymmetries. 
One of them can be used to gauge away the Goldstino, and then the resulting action coincides with that proposed by  Deser and Zumino to describe  consistent supergravity in the first-order formalism. In this sense, pure $\mathcal{N} = 1$ supergravity is a special case of the Volkov-Soroka theory. We also explain how the Volkov-Soroka approach allows one to naturally arrive at the 1.5 formalism \cite{TvN, CW}.\footnote{The terminology ``1.5 formalism'' originated from \cite{TvN}.}

Before we turn to the technical part of this paper, it is appropriate to make a few historical comments about Dmitry V. Volkov, one of the co-discoverers of supersymmetry. His most prominent results on rigid and local supersymmetry are the Goldstino model (jointly with Akulov) \cite{VA,AV} and the super-Higgs mechanism (jointly with Soroka) \cite{VS,VS2}. His approach to nonlinear realisations of internal and space-time symmetries \cite{Volkov73}, which paved the way to \cite{VA,AV,VS,VS2}, 
has also been highly influential. However, it is less known that Ref. \cite{AV} also pioneered the following fundamental concepts of modern theoretical physics: (i)  the $\cN$-extended super-Poincar\'e group and, hence,  the $\cN$-extended super-Poincar\'e algebra\footnote{The $\cN=1$ super-Poincar\'e algebra
was discovered in 1971 by Golfand and Likhtman 
\cite{GL}.} for $\cN>1$; and (ii) $\cN$-extended Minkowski superspace.
It is quite remarkable that all these results had appeared before 
the first paper by Wess and Zumino on supersymmetry \cite{WZ} was published on 18 February, 1974.\footnote{The Akulov-Volkov paper \cite{AV} was submitted to the journal {\it Theoretical and Mathematical Physics} on 8 January 1973 and published in January 1974, before the publication of the first paper on supersymmetry by Wess and Zumino \cite{WZ} and long before the work by Salam and Strathdee \cite{SS} devoted to the $\cN=1$ superspace approach. It remains largely unknown, perhaps because
it was published in a Russian journal. }

%%%%%%%%%%%%%%%%%%%%%%%%%%%%
%%%%%%%%%%%%%%%%%%%%%%%%%%%

\section{A review of the Volkov-Soroka construction} 

This section is devoted to a pedagogical review of the Volkov-Soroka construction 
 \cite{VS,VS2}.

Let  ${\mathfrak P}{(4|\cN )}$ denote the four-dimensional
$\cN$-extended super-Poincar\'e group introduced in \cite{AV}.
Any element $g \in {\mathfrak  P}{(4 | \cN )}$ is a $(4+\cN)\times (4+\cN)$ 
{supermatrix} of the form\footnote{Our parametrisation of the elements of 
${\mathfrak P}{(4|\cN )}$ follows \cite{Kuzenko2010}.}
\begin{subequations}\label{SP}
\bea
g &=& S(b,  \ve) \,h (M,U)  \equiv S h ~, 
\label{SP1} \\
S(b, \ve ) &:= &
\left(
\begin{array}{c | c ||c}
  \mathbbm{1}_2  ~& ~ 0&~{ 0}   \\
\hline
-{\rm i}\,\tilde{ b}_{(+)}  \phantom{\Big|}  ~& ~\mathbbm{1}_2 ~&~ {  2{ \e}^\dagger} \\
\hline
\hline
{ 2\e} ~& ~{ 0}&~\mathbbm{1}_\cN
\end{array}
\right)
=
\left(
\begin{array}{r | c ||c}
\d_\a{}^\b  ~& ~ 0&~{ 0}   \\
\hline
-{\rm i}\,b^{{\dot \a} \b}_{(+)}  ~& ~\d^{\dot \a}{}_{\dot \b} &~ { 2{\bar \e}^{{\dot \a} j}}
\label{SP-g} \\
\hline
\hline
{2\e_i{}^\b} ~& ~{ 0}&~\d_i{}^j
\end{array}
\right) ~, 
\label{SP2}  \\
{ h(M,U)} &:=& 
\left(
\begin{array}{c | c ||c}
 M  ~& ~ 0&~{ 0}   \\
\hline
0  ~& ~(M^{-1})^\dagger&~ { 0} \\
\hline
\hline
{ 0} ~& ~{ 0}&~U
\end{array}
\right) 
= \left(
\begin{array}{c | c| |c}
M_\a{}^\b  ~& ~ 0&~{ 0}   \\
\hline
0  ~& ({\bar M}^{-1})_{\dot \b}{}^{\dot \a}  &~ {  0} \\
\hline
\hline
{  0} ~& ~{  0}&~U_i{}^j
\end{array}
\right) ~, 
\label{SP3} 
\eea
\end{subequations}
where $M =(M_\a{}^\b)  \in \sSL(2,{\mathbb C})$, 
$U =(U_i{}^j) \in \sU(\cN)$, and 
\bea
\tilde{ b}_{(\pm)} = \tilde{ b} \pm 2\ri \e^\dagger \e~.
\label{2.2}
\eea
The tilde notation in \eqref{SP2} and \eqref{2.2} reflects the fact that there are two types of relativistic Pauli matrices, $\s_a$ and $\tilde \s_a$, see the appendix.
The group element $S(b, \ve)$
is labelled  by 
4 commuting real  parameters $b^a$ and
$4\cN $ anti-commuting
complex parameters $\ve = (\e , \e^\dagger)$, where 
$\e=( \e_i{}^\a)$  and 
$\e^\dagger =  ({\bar \e}^{{\dot \a} i} )$, 
$  {\bar \e}^{{\dot \a} i} :=
\overline{\, \e_i{}^\a \,}$. In the vector notation, eq. \eqref{2.2} reads
\bea
 b^a_{(\pm )}:= b^a \pm {\rm i} \,\e_i \s^a {\bar \e}^i 
=b^a \pm {\rm i}\, \e_i{}^\a (\s^a)_{\a \ad} {\bar \e}^{{\dot \a} i}  ~.
\label{+-}
\eea

Introduce Goldstone fields 
$Z^A (x) = \big(X^a (x) , \Q_i{}^\a (x) , \bar \Q^{\ad i} (x) \big)$
for spacetime translations $\big(X^a \big)$ and supersymmetry transformations 
$\big( \Q_i{}^\a  , \bar \Q^{\ad i}  \big)$.
They parametrise the homogeneous space
($\cN$-extended Minkowski superspace)
\bea
{\mathbb M}^{4|4\cN} 
=\frac{ {\mathfrak P}(4|\cN) }{ \sSL(2,{\mathbb C}) \times \sU(\cN) } 
\label{2.4}
\eea
according to the rule:
\bea
{\mathfrak S}(Z ) &= &
\left(
\begin{array}{c | c ||c}
  \mathbbm{1}_2  ~& ~ 0&~{ 0}   \\
\hline
-{\rm i}\,\tilde{ X}_{(+)}  \phantom{\Big|}  ~& ~\mathbbm{1}_2 ~&~ {  2{ \Q}^\dagger} \\
\hline
\hline
{ 2\Q} ~& ~{ 0}&~\mathbbm{1}_\cN
\end{array}
\right) 
~ \implies ~
{\mathfrak S}^{-1} (Z ) = 
\left(
\begin{array}{c | c ||c}
  \mathbbm{1}_2  ~& ~ 0&~{ 0}   \\
\hline
{\rm i}\,\tilde{ X}_{(-)}  \phantom{\Big|}  ~& ~\mathbbm{1}_2 ~&~  - 2{ \Q}^\dagger \\
\hline
\hline
- 2\Q ~& ~{ 0}&~\mathbbm{1}_\cN
\end{array}
\right)~,~~~
\eea
where 
\bea
\tilde{ X}_{(\pm)} = \tilde{ X} \pm 2\ri \Q^\dagger \Q~.
\eea

We define gauge super-Poincar\'e transformations by  
\bea
g(x): Z(x) \to Z'(x) ~, \qquad g {\mathfrak S}(Z) = {\mathfrak S}(Z') h~,
\eea
with $g=Sh$.
This is equivalent to the following transformations of the Goldstone fields:
\begin{subequations}\label{2.8} 
\bea
 S(b,  \ve ) :\qquad 
 \tilde X' &=& \tilde X + \tilde b +2\ri (\e^\dagger \Q- \Q^\dagger \e) ~, \label{2.8a}\\
 \Q' &=& \Q +\e \label{2.8b}
 \eea
 \end{subequations}
and 
\begin{subequations} \label{2.9}
\bea
h(M, U): \qquad \tilde{X}' &=&  (M^\dagger)^{-1} \tilde{X} M^{-1} ~,\\
\Q'&=& U \Q M^{-1} ~.
\eea
\end{subequations}

Introduce a connection $ {\mathfrak A}  = \rd x^m {\mathfrak A}_m $ taking its values in the super-Poincar\'e algebra,
\bea
{\mathfrak A}  &:= &
\left(
\begin{array}{c | c ||c}
  \O  ~& ~ 0&~{ 0}   \\
\hline
-{\rm i}\,\tilde{ e}  \phantom{\Big|}  ~& ~-\O^\dagger ~&~ {  2{ \j}^\dagger} \\
\hline
\hline
{ 2\j} ~& ~{ 0}&~\ri V
\end{array}
\right)
=
\left(
\begin{array}{r | c ||c}
\O_\a{}^\b  ~& ~ 0&~{ 0}   \\
\hline
-{\rm i}\,e^{{\dot \a} \b}  ~& ~ -\bar \O^\ad{}_\bd &~ { 2{\bar \j}^{{\dot \a} j}} \\
\hline
\hline
{2\j_i{}^\b} ~& ~{ 0}&~ \ri V_i{}^j
\end{array}
\right) ~, 
\eea
and possessing the gauge transformation law
\bea
 {\mathfrak A}' = g  {\mathfrak A} g^{-1} + g \rd g^{-1}~.
 \eea
 Here the one-forms $\O_\a{}^\b $ and $\bar \O^\ad{}_\bd $ are the spinor counterparts of the Lorentz connection $\O^{ab} = \rd x^m \O_m{}^{ab} = - \O^{ba} $ such that 
 \bea
 \O_\a{}^\b = \hf (\s^{ab})_\a{}^\b\, \O_{ab} ~, \qquad 
\bar  \O^\ad{}_\bd =- \hf (\tilde \s^{ab})^\ad{}_\bd \,\O_{ab} ~.
 \eea
 The Lorentz connection is an independent field. It is expressed in terms of the other fields on the mass shell. 
 The one-form $e^{{\dot \a} \b}$ is the spinor counterpart of the vierbein 
 $e^a = \rd x^m e_m{}^a$.  The fermionic one-forms $\j_i{}^\b $
 and ${\bar \j}^{ \ad j} $ describe $\cN$ gravitini. Finally, the anti-Hermitian one-form 
 $\ri V$ is the $\sU(\cN)$ gauge field. 
 
Associated with ${\mathfrak S}$ and ${\mathfrak A}$ is another connection 
\bea
{\mathbb A} := {\mathfrak S}^{-1}{\mathfrak A} {\mathfrak S} 
+ {\mathfrak S}^{-1} \rd {\mathfrak S} 
~, 
\eea
which is characterised by the gauge transformation law
\bea
{\mathbb A}' = h {\mathbb A} h^{-1} + h \,\rd h^{-1}~,
\label{2.14}
\eea 
for an arbitrary gauge parameter $g = Sh$.
This transformation law tells us that $\mathbb A$ is invariant under all gauge 
transformations of the form $g = S(b, \ve)$ which describe local spacetime translations and supersymmetry transformations.  
The connection ${\mathbb A} $ is the main object of the Volkov-Soroka construction. 
It has the form 
\bea
{\mathbb A}  &:= &
\left(
\begin{array}{c | c ||c}
  \O  ~& ~ 0&~{ 0}   \\
\hline
-{\rm i}\,\tilde{ E}  \phantom{\Big|}  ~& ~-\O^\dagger ~&~ {  2{ \J}^\dagger} \\
\hline
\hline
{ 2\J} ~& ~{ 0}&~ \ri V
\end{array}
\right)~,
\eea
where we have defined 
\begin{subequations}\label{2.15}
\bea
\J &:=& \j +\cD \Q~,  \label{2.15a} \\
\J^\dagger &:=& \j^\dagger +\cD \Q^\dagger~, 
\label{2.15.b} \\
\tilde{ E} &:=& \tilde{ e} + \cD \tilde{ X} + 4\ri (\J^\dagger - \hf \cD \Q^\dagger) \Q 
- 4\ri \Q^\dagger ( \J -\hf \cD \Q)~, 
\label{2.15c}
\eea
\end{subequations}
and $\cD$ denotes the covariant derivative, 
\begin{subequations}
\bea
\cD \Q &=& \rd \Q - \Q \O + \ri V \Q~, \\
\cD \Q^\dagger &=& \rd \Q^\dagger - \O^\dagger \Q^\dagger - \ri  \Q^\dagger V ~,\\
\cD \tilde{ X} &=& \rd  \tilde{ X} -\O^\dagger  \tilde{ X} -  \tilde{ X} \O~.
 \eea
 \end{subequations}

Equation \eqref{2.14} is equivalent to the following gauge transformation laws:
\begin{subequations}
\bea
\O' &=&  M \O M^{-1}  +M \rd M^{-1} ~,\\
\ri V'&=& U (\ri V )U^{-1}  +  U \rd U^{-1} 
\eea
\end{subequations}
and 
\begin{subequations}
\bea
\tilde{E}' &=&  (M^\dagger)^{-1} \tilde{E} M^{-1} ~,\\
\J'&=& U \J M^{-1} ~, 
\eea
\end{subequations}
We see that the supersymmetric one-forms $E^a$ and $\J_i{}^\b$ transform as tensors with respect to the Lorentz and $\sU(\cN)$ gauge groups.

Making use of \eqref{2.8}, we deduce the local supersymmetry transformation
of the gravitini and the vielbein
\begin{subequations}
\bea
\j' &=& \j - \cD \e ~,\\
\tilde{e}' &=&  \tilde e + 4 \ri (\e^\dagger \j - \j^\dagger \e) 
+2\ri (\cD \e^\dagger \e - \e^\dagger \cD \e) ~.
\eea
\end{subequations} 
In the infinitesimal case, this transformation can be rewritten in the form 
\bea
\d_\ve \j = - \cD \e~, \qquad 
\d_\ve e^a =  2\ri \, {\rm tr} \big[ \s^a (\j^\dagger \e  - \e^\dagger  \j ) \big] ~.
\label{2.21}
\eea
As pointed out by Volkov \cite{Volkov1}, the transformation laws in \eqref{2.21} coincide with those used by Deser and Zumino in their construction of $\cN=1$ supergravity \cite{DZ}.
We should remark that the supersymmetry transformations of the Goldstone fields $X^a$ and $\Q_i{}^\b$ are given by the relations \eqref{2.8}.

Let us consider a local Poincar\'e translation, $S(b,0)$. It only acts on the Goldstone vector field $X^a$ and the vierbein $e^a$, 
\bea
X'{}^a = X^a + b^a~, \qquad e'{}^a = e^a - \cD b^a~.
\label{2.22}
\eea
We have two types of gauge transformations with vector-like parameters, the general coordinates transformations and the local Poincar\'e translations. The latter gauge freedom can be fixed by imposing the condition $X^a =0$, and then we stay only with the general coordinate invariance. However, in what follows we will keep $X^a$ intact. 

The curvature tensor is given by 
\bea
{\mathbb R}=\rd {\mathbb A} - {\mathbb A} \wedge {\mathbb A}~, \qquad 
{\mathbb R}' = h {\mathbb R} h^{-1}~.
\eea
Its explicit form is 
\bea
{\mathbb R}  &:= &
\left(
\begin{array}{c | c ||c}
  R  ~& ~ 0&~{ 0}   \\
\hline
-{\rm i}\,\tilde{ \mathbb T}  \phantom{\Big|}  ~& ~-R^\dagger ~&~   2 \cD \J^\dagger \\
\hline
\hline
 2\cD \J ~& ~{ 0}&~ \ri F
\end{array}
\right)~,
\eea
where $R = (R_\a{}^\b) $ and $R^\dagger = (\bar R^\ad{}_\bd )$
form the Lorentz curvature, $F=(F_i{}^j)$ is the 
Yang-Mills  field strength,
\begin{subequations}
\bea
\cD \J &=& \rd \J - \J \wedge \O - \ri V \wedge \J~,
\\
\cD \J^\dagger &=& \rd \J^\dagger +\O^\dagger \wedge \J^\dagger 
- \ri \J^\dagger \wedge V
\eea
\end{subequations} 
are the gravitino field strengths, and 
\bea
\tilde {\mathbb T} = \rd \tilde E - \tilde E \wedge \O + \O^\dagger \wedge \tilde E 
- 4\ri \J^\dagger \wedge \J = \cD \tilde E - 4\ri \J^\dagger \wedge \J 
\eea
is the supersymmetric torsion tensor.  In the vector notation, the torsion tensor reads
\bea
{\mathbb T}^a = \cD E^a + 2\ri \J \wedge \s^a \bar \J~.
\label{torsion}
\eea

It should be pointed out that the exterior derivative is defined to obey the property 
\bea
\rd \Big( \S_p \wedge \S_q \Big) = \S_p \wedge \rd \S_q 
+(-1)^q \rd \S_p \wedge \S_q~,
\eea
which is used for superforms \cite{WB}.

The above results allow one to engineer gauge-invariant functionals that can be used to construct a locally supersymmetric action. The invariants proposed in \cite{VS,VS2} are the following:
\begin{itemize}
\item The Einstein-Hilbert action 
\bea
S_{\rm EH} = \frac 14 \int \ve_{abcd} E^a \wedge E^b \wedge R^{cd} ~;
\label{EH}
\eea
\item The Rarita-Schwinger action 
\bea
S_{\rm RS} = \hf \int \Big( \J_i \wedge E^a \wedge \s_a \cD \bar \J^i - \cD \J_i \wedge E^a \wedge \s_a \bar \J^i \Big) ~;
\label{RS}
\eea
\item The cosmological term
\bea
S_{\rm cosmological } = \frac{1}{24} \int \ve_{abcd} E^a \wedge E^b \wedge E^{c} \wedge E^d ~.
\label{2.31}
\eea
\item $\sO(\cN)$-invariant mass term 
\bea
S_{\rm mass} = \frac{\ri}{4}  \int  E^a \wedge E^b \wedge 
\Big(\d^{ij}  \J_i \wedge \s_{ab} \J_j  - \d_{ij} \bar \J ^i\wedge \tilde{\s}_{ab} \bar \J^j \Big) ~.
\label{2.32}
\eea
\end{itemize}
The functionals $S_{\rm EH} $, $S_{\rm RS} $ and $S_{\rm cosmological}$ are $\sU(\cN)$ invariant.  The cosmological term, eq. \eqref{2.31},  also  contains the kinetic term for the Goldstini \cite{VA,AV}.
The mass term \eqref{2.32} is invariant under local internal transformations only if the group $\sU(\cN)$ is replaced with $\sO(\cN)$, and the gauge connection $\ri V$ takes its values in the Lie algebra $\mathfrak{so}(\cN)$. The Yang-Mills action associated with $V$ is obviously supersymmetric, but it will not be used  in what follows. 

In the  $\cN=1$ case, the Volkov-Soroka theory is described by the general action 
\bea
S= S_{\rm EH} +4c \,S_{\rm RS} + 4m\, S_{\rm mass} + \l S_{\rm cosmological} ~,
\label{2.33}
\eea
with $c$, $m$ and $\l$ coupling constants.

%%%%%%%%%%%%%%%%%%%%%%%%%%%%%
%%%%%%%%%%%%%%%%%%%%%%%%%%%%%

\section{Second local supersymmetry} \label{section3}

In this section our consideration is restricted to the $\cN=1$ case, and  the $\sU(1)$ 
gauge field is switched off, $V=0$. We are going to  demonstrate that the action 
\bea
S_{\rm SUGRA} = S_{\rm EH} + 4 S_{\rm RS} 
\label{3.1}
\eea
is invariant under 
a new local supersymmetry transformation described by the parameter  ${\bm \ve} = ({\bm \e}^\a , \bar{\bm \e}^\ad)$.
It acts on the composite fields 
$\J^\a$ and $E^a$, defined by eq. \eqref{2.15}, and the Lorentz connection as follows
\begin{subequations} \label{3.2}
\bea
\d_{\bm \ve} \J^\a &=& - \cD {\bm \e}^\a~, \qquad 
\d_{\bm \ve} E^a =  2\ri \, (\J \s^a \bar {\bm \e}   - {\bm \e} \s^a \bar \J )  ~, \label{3.2a} \\
  \frac 14  \ve_{abcd} \d_{\bm \ve} \O^{bc} \wedge E^d &=& 
  {\bm \e} \s_a \cD \bar \J 
  +  \cD \J \s_a \bar{\bm \e} ~. \label{3.2b}
\eea
The Goldstone fields are inert under this transformation,
\bea
\d_{{\bm \ve}}  X^a &=& 0~, \qquad \qquad \d_{ {\bm \ve}} \Q^\a =0~. \label{3.2c}
\eea
The elementary field $\j^\a $ and $e^a$ transform as follows:
\bea
\d_{\bm \ve} \j^\a &=& - \cD {\bm \e}^\a +\Q^\b \d_{\bm \ve} \O_\b{}^\a~, \label{3.2d}\\
\d_{\bm \ve} e^a &=&  2\ri \big( \J \s^a \bar {\bm \e} - {\bm \e} \s^a \bar \J \big)
 +2\ri \big( \Q \s^a \cD \bar {\bm \e} - \cD {\bm \e} \s^a \bar \Q \big) 
 -\d_{\bm \ve} \O^a{}_b X^b\non \\
 && + \hf \ve^{abcd} \d_{\bm \ve} \O_{bc} \Q\s_d \bar \Q~.
 \label{3.2e}
\eea
\end{subequations}
It should be pointed out that the transformation laws in \eqref{3.2a} can be viewed as  a natural generalisation of the Volkov-Soroka local supersymmetry \eqref{2.21}.

The dependence on $\d_{\bm \ve} \O$ in \eqref{3.2d} and \eqref{3.2e} is such that  
the composite fields $\J^\a$ and $E^a$ remain unchanged when the connection gets the displacement $\O \to \O + \d_{\bm \ve} \O$. We should point out  that the action \eqref{3.1} 
involves only  the one-forms  $\J^\a$, $\bar \J^\ad$, $E^a$ and $\O^{ab}$ and their descendants. We also remark that the transformations \eqref{3.2a} and \eqref{3.2b} reduce to those given by Deser and Zumino \cite{DZ} if the Goldstone fields 
$X^a$ and $\Q^\a$ are switched off.

It should be pointed out that eq. \eqref{3.2b} uniquely determines 
$\d_{\bm \ve} \O^{bc}$. Indeed, given a vector-valued two-form
\bea
\S_a = \hf E^c \wedge E^b \S_{a,bc} ~,
\eea
the following equation 
\bea
\hf \ve_{abcd}  \o^{bc} \wedge E^d \equiv \widetilde{\o}_{ab} \wedge E^b = \S_a~,
\qquad  \widetilde{\o}_{ab} = E^c \widetilde{\o}_{c,ab} 
\eea
on the one-form ${\o}^{ab} = E^c {\o}_c{}^{ab} $, which takes its values in the Lorentz algebra,  
has the unique solution 
\bea
\widetilde{\o}_{c,ab} = \hf \Big( \S_{a,bc} - \S_{b,ac} - \S_{c,ab} \Big)~.
\eea

Now we turn to demonstrating that the action \eqref{3.1} is invariant under the supersymmetry transformation \eqref{3.2}. Let us first take into account the variations 
in \eqref{3.2a}.
Varying the Einstein-Hilbert action \eqref{EH} for $\bm \e \neq 0$ and $\bar {\bm \e} =0$
gives
\bea
\d^{(1)}_{\bm \e} S_{\rm EH} = -\ri  \int \ve_{abcd} 
R^{ab} \wedge E^c \wedge {\bm \e} \s^d \bar \J   ~.
\label{3.3}
\eea
Varying the Rarita-Schwinger action \eqref{RS} gives
\bea
\d^{(1)}_{\bm \e} S_{\rm RS} &=& - \int \Big\{ 
\hf  \cD  E^a \wedge {\bm \e} \s_a \cD \bar \J 
-\ri {\bm \e}\s^a \bar \J \wedge \J \wedge \s_a \cD \bar \J
\Big\}
 \non \\
&&
+\hf \int \Big\{ E^a \wedge {\bm \e} \s_a R^\dagger \wedge  \bar \J  
-E^a \wedge {\bm \e} R \wedge \s_a \bar \J
\Big\} ~,
\label{3.4}
\eea
where we have used the relations 
\bea
\cD \cD {\bm \e} = - {\bm \e} R~, \qquad 
\cD \cD \J^\dagger = - R^\dagger \wedge \J^\dagger~.
\eea
Here $\d^{(1)}$ means that we have taken into account only the variations \eqref{3.2a}.

Making use of the identities \eqref{A.14}, the curvature-dependent contributions in \eqref{3.4} can be rearranged as
\bea
\hf \int \Big\{ E^a \wedge {\bm \e} \s_a R^\dagger \wedge \bar \J  
-E^a \wedge {\bm \e} R \wedge \s_a \bar \J
\Big\} =
\frac{\ri}{4}  \int \ve_{abcd} 
R^{ab} \wedge E^c \wedge {\bm \e} \s^d \bar \J  ~.
\label{3.6}
\eea
In order for the curvature contributions in \eqref{3.3} and \eqref{3.4}  to cancel each other, we must consider the action \eqref{3.1}, for which we obtain
\bea
\d^{(1)}_{\bm \e} (S_{\rm EH} + 4 S_{\rm RS} ) 
=  -2 \int \Big( 
  \cD  E^a + 2\ri \J \wedge \s^a \bar \J \Big) 
  \wedge {\bm \e} \s_a \cD \bar \J 
  =-2 \int  {\bm \e} \s_a \cD \bar \J \wedge {\mathbb T}^a 
  ~,~~~
  \eea
with ${\mathbb T}^a$ being the supersymmetric torsion tensor
\eqref{torsion}. Adding the complex conjugate part gives
\bea
\d^{(1)}_{\bm \ve} (S_{\rm EH} + 4 S_{\rm RS} ) 
  = -2\int \Big(  {\bm \e} \s_a \cD \bar \J 
  +  \cD \J \s_a \bar{\bm \e} \Big)\wedge {\mathbb T}^a 
  ~.
  \label{3.8}
  \eea

%%%%%%%%%%%%%%%%%%%%%%%%%%%%%%%%%%%

Next, let us vary the action \eqref{3.1} with respect to the Lorentz connection $\O^{ab}$.
We give the Lorentz connection a small disturbance, $ \O \to \O + \d_{\bm \ve}  \O $, 
with $  \d_{\bm \ve} \O $ to be determined below, and assume that the fields  $\j^\a$ and $e^a$ also acquire $  \d_{\bm \ve} \O $-dependent variations given in \eqref{3.2d} and \eqref{3.2e}. 
We denote $\d^{(2)}$ the corresponding variation. 
Direct calculations give 
\bea
\d^{(2)}_{\bm \ve} (S_{\rm EH} + 4 S_{\rm RS} ) 
=  \hf \int \ve_{abcd} \d_{\bm \ve} \O^{bc} \wedge E^d \wedge {\mathbb T}^a~.
\label{3.9}
\eea

Combining the results \eqref{3.8} and \eqref{3.9}, we end up with  
\bea
\d_{\bm \ve}  (S_{\rm EH} + 4 S_{\rm RS} ) 
=-2\int \Big(  {\bm \e} \s_a \cD \bar \J 
  +  \cD \J \s_a \bar{\bm \e} 
  -\frac 14  \ve_{abcd} \d_{\bm \ve} \O^{bc} \wedge E^d 
  \Big)\wedge {\mathbb T}^a ~.
\eea
This variation vanishes if $\d_{\bm \ve} \O$ is given by eq. \eqref{3.2b}.

We have demonstrated that the theory \eqref{3.1} has two types of local supersymmetry. 
The original Volkov-Soroka supersymmetry is described by the relations 
\begin{subequations} \label{3.11}
\bea
\d_{ \ve} \j^\a &=& - \cD { \e}^\a~, \qquad 
\d_{\ve} e^a =  2\ri \, (\j \s^a \bar { \e}   - { \e} \s^a \bar  \j )  ~, \\
\d_{ { \ve}} \Q^\a &=&\e^\a ~,\qquad
\d_{{ \ve}}  X^a = \ri (\Q \s^a \bar \e-  \e \s^a \bar \Q)~.
\eea
\end{subequations}
The second supersymmetry, introduced in this work, 
 is given by equation \eqref{3.2}. The gauge transformations
\eqref{2.22} and  \eqref{3.11} allow us to gauge away the Goldstone fields 
$Z^A (x) = \big(X^a (x) , \Q^\a (x) , \bar \Q^{\ad } (x) \big)$
by 
imposing the conditions 
\bea
X^a=0~, \qquad 
\Q^\a=0~.
\label{3.15}
\eea
As a result, the action \eqref{3.1} turns into the supergravity action proposed by Deser and Zumino \cite{DZ}, and the local supersymmetry transformations \eqref{3.2a} turn into those given in \cite{DZ}.

%%%%%%%%%%%%%%%%%%%%%%%%%%%%%%%%%%
%%%%%%%%%%%%%%%%%%%%%%%%%%%%%%%%%%

\section{Conclusion} 

 A few years ago, it was shown\footnote{Similar conclusions had been obtained earlier in \cite{IvanovK}.}  \cite{BMST} that the Volkov-Soroka theory \eqref{2.33},
with non-vanishing parameters $c$, $m$ and $\l$,
is equivalent to spontaneously broken $\cN=1$ supergravity \cite{DZ2,LR}
which was called de Sitter supergravity in \cite{BFKVP}. 

In this paper we have demonstrated that the action \eqref{3.1} is a St\"uckelberg-type extension of the $\cN=1$ supergravity theory in the first-order formalism proposed by Deser and Zumino.
Equivalently, the latter theory is a gauged-fixed version of \eqref{3.1}.
Therefore,  pure $\mathcal{N} = 1$ supergravity is a special case of the Volkov-Soroka theory.

The new local supersymmetry \eqref{3.2} of the action \eqref{3.1} is the main original result of this paper. The Goldstino is just a compensator for the first local supersymmetry in this theory. In the gauge \eqref{3.15} the action \eqref{3.1} turns into the supergravity action in the first-order formalism.  

In the above analysis, the Lorentz connection $\O$ was an independent field. 
The alternative approach is to work with a composite connection obtained 
  by imposing the constraint 
\bea
{\mathbb T}^a = \cD E^a + 2\ri \J \wedge \s^a \bar \J=0~,
\label{4.1}
\eea
which is an example of the so-called inverse Higgs mechanism \cite{IO}.
The constrain \eqref{4.1} is invariant under the first supersymmetry transformation.
 In the gauge \eqref{3.15}, the constraint \eqref{4.1} is uniquely solved to give the standard expression for the connection, $\O = \O(e,\j)$, in terms of the vielbein and the gravitino,
 see e.g. \cite{VanNieuwenhuizen:1981ae} for a review.\footnote{For non-vanishing Goldstone fields $X^a$ and $\Q^\a$,  the constraint \eqref{4.1} should also allow one to uniquely determine the connection in terms of the other fields. But in this case the equation \eqref{4.1} becomes highly nonlinear, and its explicit solution is hard to derive.}
 Under the second supersymmetry transformation, the connection varies in such a way that \eqref{4.1} is preserved. It follows from the relations \eqref{3.8} and \eqref{3.9} that the action \eqref{3.1} is invariant under the second supersymmetry transformation. We observe that the Volkov-Soroka approach allows one to naturally arrive at the 1.5 formalism 
 \cite{TvN, CW}.\footnote{Our consideration clearly shows that the work by Volkov and Soroka \cite{VS2} contained all prerequisites that could, in principle, be used to discover supergravity before 1976. It is natural to wonder why they did not discover supergravity. Of course, they did not ask the right question in \cite{VS,VS2}. It seems more important, however, that their ideas were well ahead of time, and the scientific community in the Soviet Union  was not ready to accept the novel concepts put forward in these publications. This is similar to the discovery of rigid supersymmetry in four dimensions by Golfand and Likhtman \cite{GL} (see \cite{Shifman} for a historical account) whose work was not appreciated in the Soviet Union.}
 
The novelty of our work is that we have developed a new nonlinear realisation approach to constructing unbroken simple Poincar\'e supergravity theories.\footnote{In the framework of the Ogievetsky-Sokatchev approach \cite{OS} to the old minimal formulation for $\cN=1$ supergravity \cite{WZ78,old1,old2}, Ivanov and Niederle \cite{IN} described $\cN=1$ supergravity as a nonlinear realisation. Their construction is completely different from ours.} 
We only studied the case of $\cN=1$ supergravity in four dimensions, but it seems that the same approach can be used in other dimensions.
 Specifically, as in \cite{VS,VS2} one introduces Goldstone fields $Z^A (x) = \big(X^a (x) , \Q^{\hat \a} (x)  (x) \big)$
for spacetime translations $\big(X^a \big)$ and supersymmetry transformations 
$\big( \Q^{\hat \a} \big)$, with $\hat \a$ denoting a spinor index. These fields parametrise the coset space, that is Minkowski superspace.
To describe unbroken supergravity, the Goldstone fields must describe compensating degrees of freedom. This means there should be two types of gauge transformations with vector parameters, and also two types of local supersymmetry transformations, in order to be able to gauge away the Goldstone fields. By construction, there are always two types of gauge transformations with vector parameters, the general coordinates transformations and the local Poincar\'e translations. The latter gauge freedom can be fixed by imposing the condition $X^a =0$, and then we stay only with the general coordinate invariance. By construction, there is always one type of local supersymmetry. A second local supersymmetry emerges only for a special choice of the parameter in the action.\footnote{It is worth pointing out that similar ideas have been used to describe AdS gravity \cite{West,SW1,SW2} and AdS supergravity \cite{PV}.} The described approach can definitely be used to provide a new derivation of $\cN=1$ topologically massive supergravity in three dimensions originally constructed in \cite{DK}.

As pointed out earlier, the model \eqref{3.1} is a St\"uckelberg reformulation of the unbroken $\cN=1$ supergravity in the first-order formalism. It is known that the St\"uckelberg formalism is often useful in the quantum theory. It would interesting to revisit the quantisation of $\cN=1$ supergravity using the novel formulation \eqref{3.1}.
\\

\noindent
{\bf Acknowledgements:}\\ 
I thank Evgeny Buchbinder, Ian McArthur,  Dmitri Sorokin, and especially Paul Townsend and Arkady Tseytlin for discussions, and Darren Grasso and Michael Ponds for comments on the manuscript.
I also gratefully acknowledge email correspondence with Ioseph Buchbinder, Stepan Douplii, Jim Gates, Evgeny Ivanov, Ulf Lindstr\"om and Martin Ro\v{c}ek.
My special thanks go to Jake Stirling for checking the calculations and correcting typos in section 
\ref{section3}.
This work was supported in part by the Australian Government through
the Australian Research Council, project No. DP200101944.

\appendix 

\section{Two-component spinor formalism}\label{appendixA}

In this appendix we collect the key formulae of the two-component spinor formalism.
Our notation and two-component spinor conventions correspond to those used in 
\cite{WB,BK}. In particular, the Minkowski metric is $\eta_{ab}= {\rm diag} \,(-1, +1, +1, +1)$, and the Levi-Civita tensor $\ve^{abcd}$ is normalised by $\ve^{0123} =1$.

Given a four-vector $p^a$, it can equivalently be described 
as an Hermitian $2\times2$ matrix with lower spinor indices
\bea
{p}: = p^a \s_a = p^\dagger =( p_{\a  \bd})~, 
\qquad { \s_a = ({\mathbbm 1}_2, \vec{\s} )}~,
\eea
or as an Hermitian $2\times2$ matrix with upper spinor indices
\bea
\tilde{ p}: = p^a \tilde{\s}_a =\tilde{p}{}^\dagger =({ p^{ \ad \b}})~, 
\qquad { \tilde{\s}_a = ({\mathbbm 1}_2, -\vec{\s} )}~, 
\eea
with $ \vec \s$ being the Pauli matrices. The two sets of the relativistic Pauli matrices, 
$\s_a$ and $\tilde{\s}_a$, are related to each other by the rule
\bea
(\tilde \s_a)^{\ad \a} = \ve^{\ad \bd} \ve^{\a\b} (\s_a)_{\b \bd} ~,
\eea
where $ \ve^{\a \b} $ and $ { \ve_{\a \b}} $, $ \ve^{\ad \bd} $ and $ { \ve_{\ad \bd}} $
are antisymmetric spinor metrics normalised as
$\ve^{12}= \ve_{21} =1$
and
$\ve^{\dot 1 \dot 2}= \ve_{\dot 2 \dot1} =1$.
These are used to raise and lower the spinor indices, 
\bea
 \j^\a =  \ve^{\a \b} \,\j_\b~, \qquad \j_\a =  \ve_{\a \b} \,\j^\b ~,
\eea
and similarly for the dotted spinors.

Let $ {\mathfrak P}(4)$ be the universal covering group of 
the restricted Poincar\'e group $ {\sISO}_0(3,1)$.
It is usually realised as the group of linear inhomogeneous transformations 
$(M, b)$ acting 
on the space of $2\times 2$ Hermitian matrices 
$ x = x^a \s_a =x^\dagger $ 
as follows:
\bea
x \to x' =  {x'}{}^a \s_a =MxM^\dagger + b ~, 
\quad b = b^a \s_a = b^\dagger~, \quad 
M = (M_\a{}^\b ) \in {\sSL}(2,{\mathbb C})~.
\label{A.6}
\eea
Here $M^\dagger := {\bar M}^{\rm T}$ is the Hermitian conjugate of $M$,
and $\bar M =({\bar M}_{\ad}{}^{\bd} )$ the complex conjugate of $M$, 
with ${\bar M}_{\ad}{}^{\bd} := \overline{ M_\a{}^\b}$. The group   $ {\mathfrak P}(4)$ is equivalently realised  
as the group of linear inhomogeneous transformations acting 
  on the space of $2\times 2$ Hermitian matrices 
$\tilde{x}: = x^a \tilde{\s}_a =\tilde{x}{}^\dagger$
as follows:
\bea
\tilde{ x} \to \tilde{x}' =  x'^a \tilde{\s}_a =(M^{-1})^\dagger \tilde{ x}M^{-1} + \tilde{ b} ~,
\quad \tilde{ b} = b^a \tilde{\s}_a ~.
\label{A.7}
\eea
In Minkowski space  $\mathbb{M}^4\equiv {\mathbb R}^{3,1}$, the transformation 
\eqref{A.6} or, equivalently,  \eqref{A.7} looks like 
\bea
x'^a = \big( \L (M) \big)^a{}_b \,x^b+b^a~, \qquad
\big( \L (M) \big)^a{}_b = -\hf \tr \big( \tilde{\s}{}^a M \s_b M^\dagger\big)~.
\eea

It is also possible to realise ${\mathfrak P}(4)$ as a subgroup of $\sSU(2,2)$
consisting of all block triangular matrices of the form:
\bea
&& { (M, {b})} := \left(
\begin{array}{c | c}
  M  ~& ~ 0   \\
\hline
-{\rm i}\,\tilde{ b} \,M 
\phantom{\Big(}
& ~(M^{-1})^\dagger\\
\end{array}
\right) = { (\mathbbm{1}_2, {b}) \, (M,0 )}~,
\eea
with $M$ and $\tilde b$ as in \eqref{A.6} and \eqref{A.7}. 
Minkowski space is the homogeneous space 
\bea
{\mathbb M}^4= {\mathfrak P}(4)/ {\sSL}(2,{\mathbb C})~,
\eea 
compare with eq. \eqref{2.4} defining the $\cN$-extended Minkowski superspace.
Its points are naturally  parametrised by the Cartesian coordinates 
$x^a $ 
corresponding to the coset representative:
\bea
(\mathbbm{1}_2, x) = 
\left(
\begin{array}{r | c}
  \mathbbm{1}_2  ~& ~ 0   \\
\hline
-{\rm i}\,\tilde{ x}  ~& ~\mathbbm{1}_2\\
\end{array}
\right) ~.
\eea

Given an antisymmetric tensor field $F_{ab} = - F_{ba}$,
it can be equivalently described by a symmetric rank-two spinor 
$F_{\a\b} = F_{\b\a}$ and its conjugate $\bar F_{\ad\bd}$.
The precise correspondence $F_{ab} \longleftrightarrow (F_{\a\b}  , \bar F_{\ad\bd})$ 
is given by 
\bea 
F_{ab} = (\s_{ab})^{\a\b} F_{\a\b} - (\tilde{\s}_{ab})^{\ad\bd} \bar F_{\ad\bd} \ , \quad 
F_\a{}^\b := \hf (\s^{ab})_\a{}^\b F_{ab} \ , 
\quad \bar F^\ad{}_\bd := - \hf (\tilde{\s}^{ab})^\ad{}_\bd F_{ab} ~.\non
\eea
Here the matrices $\s_{ab} = \Big( (\s_{ab})_\a{}^\b\Big)$ and 
$\tilde \s_{ab} = \Big( (\tilde \s_{ab})^\ad{}_\bd\Big)$ are defined by 
\bea
\s_{ab} = -\frac 14 (\s_a \tilde \s_b - \s_b \tilde \s_a) ~, \qquad 
\tilde \s_{ab} = -\frac 14 (\tilde \s_a  \s_b - \tilde \s_b  \s_a) ~.
\eea
These matrices are (anti) self-dual, 
\bea
\hf \ve^{abcd} \s_{cd} = -\ri \s^{ab} ~, \qquad 
\hf \ve^{abcd} \tilde \s_{cd} = \ri \tilde \s^{ab} ~.
\eea
The important identities involving $\s_{ab}$ and $\tilde \s_{ab}$ are:
\begin{subequations}\label{A.14}
\bea
\s_{ab} \s_c &=& - \hf \big( \eta_{ac} \s_b - \eta_{bc} \s_a\big) 
- \frac{\ri}{2} \ve_{abcd} \s^d~, \\
\s_c \tilde \s_{ab} &=&\phantom{-}  \hf \big( \eta_{ac} \s_b - \eta_{bc} \s_a\big) 
- \frac{\ri}{2} \ve_{abcd} \s^d~.
\eea
\end{subequations} 

%%%%%%%%%%%%%%%%%%%%%%%%%%%%%%%%%%%
%%%%%%%%%%%%%%%%%%%%%%%%%%%%%%%%%%%

\begin{footnotesize}

\end{footnotesize}

%%%%%%%%%%%%%%%%%%%%%%%%%%%%%%%%%%%%
%%%%%%%%%%%%%%%%%%%%%%%%%%%%%%%%%%%%


\begin{thebibliography}{66}




\bibitem{FvNF}
D.~Z.~Freedman, P.~van Nieuwenhuizen and S.~Ferrara,
``Progress toward a theory of supergravity,''
Phys. Rev. D \textbf{13}, 3214 (1976).
%doi:10.1103/PhysRevD.13.3214

%\cite{Deser:1976eh}
\bibitem{DZ}
S.~Deser and B.~Zumino,
``Consistent supergravity,''
Phys. Lett. B \textbf{62}, 335 (1976).
%doi:10.1016/0370-2693(76)90089-7


\bibitem{Volkov1}
D.~V.~Volkov,
``Supergravity before and after 1976,'' in: {\it 
Concise Encyclopedia of Supersymmetry and Noncommutative Structures in Mathematics and Physics}, S.~Duplij, W.~Siegel and J.~Bagger (Eds.),
Kluwer Academic Publishers (2004), pp. 6--9
[arXiv:hep-th/9404153 [hep-th]].

\bibitem{Volkov2}
D.~V.~Volkov,
``Supergravity before 1976,'' in: 
{\it History of Original Ideas and Basic Discoveries in Particle Physics}, 
H. B. Newman and T. Ypsilantis (Eds.), 
Plenum Press, New York (1996), pp. 663-675
%[NATO Sci. Ser. B \textbf{352},  663(1996)],
%doi:10.1007/978-1-4613-1147-8\_34
[arXiv:hep-th/9410024 [hep-th]].

\bibitem{VS} 
  D.~V.~Volkov and V.~A.~Soroka,
  ``Higgs effect for Goldstone particles with spin 1/2,''
  JETP Lett.\  {\bf 18}, 312 (1973)
  [Pisma Zh.\ Eksp.\ Teor.\ Fiz.\  {\bf 18}, 529 (1973)].

\bibitem{VS2}
D.~V. Volkov and V.~A. Soroka, ``Gauge fields for symmetry group with
  spinor parameters,''   Theor. Math. Phys. {\bf 20}, 829 (1974)  
[Teor. Mat. Fiz. {\bf 20}, 291(1974)].

\bibitem{Soroka1}
V.~A.~Soroka,
``Starting point of supergravity,''
[arXiv:hep-th/0111271 [hep-th]].

\bibitem{Soroka2}
V.~A.~Soroka,
``The Sources of supergravity,''
in {\it The supersymmetric World: The beginning of the Theory},
G. Kane and M. Shifman (Eds.), World Scientific, Singapore (2000),
pp. 88--92
[arXiv:hep-th/0203171 [hep-th]].
%0 citations counted in INSPIRE as o

\bibitem{Nurmagambetov}
A.~J.~Nurmagambetov,
``How old supergravity is: Thirty five years or more?,''
Prob. Atomic Sci. Technol. \textbf{2011N5}, 3 (2011).

\bibitem{Dup}
S.~Duplij,
``Supergravity was discovered by D.V. Volkov and V.A. Soroka in 1973, wasn't it?,''
Eur. J. Phys. \textbf{3}, 81 (2019)
%doi:10.26565/2312-4334-2019-3-10
[arXiv:1910.03259 [physics.hist-ph]].

\bibitem{TvN}
P.~K.~Townsend and P.~van Nieuwenhuizen,
``Geometrical interpretation of extended supergravity,''
Phys. Lett. B \textbf{67}, 439 (1977).
%doi:10.1016/0370-2693(77)90439-7

\bibitem{CW}
A.~H.~Chamseddine and P.~C.~West,
``Supergravity as a gauge theory of supersymmetry,''
Nucl. Phys. B \textbf{129}, 39 (1977).

\bibitem{VA}
D.~V.~Volkov and V.~P.~Akulov,
``Possible universal neutrino interaction,''
  {JETP Lett.\  {\bf 16}, 438 (1972)}   
  [Pisma Zh.\ Eksp.\ Teor.\ Fiz.\   {\bf 16},  621 (1972)]; 
  ``Is the neutrino a Goldstone particle?,''
  Phys.\ Lett.\  B {\bf 46}, 109 (1973).

\bibitem{AV}
V.~P. Akulov and D.~V. Volkov, ``Goldstone fields with spin 1/2,''
   Theor. Math. Phys. {\bf 18}, 28 (1974)  28 [Teor. Mat. Fiz. {\bf 18}, 39 (1974)].
 
\bibitem{Volkov73}
D.~V.~Volkov, ``Phenomenological Lagrangians,''
Sov. J. Particles Nucl. {\bf 4}, 1 (1973) 
[Fiz. Elem. Chast. Atom. Yadra \textbf{4}, 3 (1973)].

%\cite{Golfand:1971iw}
\bibitem{GL}
Yu.~A.~Golfand and E.~P.~Likhtman,
``Extension of the algebra of Poincar\'e group generators and violation of P  invariance,''
  JETP Lett.\  {\bf 13}, 323 (1971)  [Pisma Zh.\ Eksp.\ Teor.\ Fiz.\  {\bf 13}, 452 (1971)].
  %%CITATION = ZFPRA,13,452;%%

%\cite{Wess:1974tw}
\bibitem{WZ}
J.~Wess and B.~Zumino,
``Supergauge transformations in four dimensions,''
Nucl.\ Phys.\  B {\bf 70}, 39 (1974).
  %%CITATION = NUPHA,B70,39;%%


%\cite{Salam:1974yz}
\bibitem{SS}
  A.~Salam and J.~A.~Strathdee,
  ``Super-gauge transformations,''
 Nucl.\ Phys.\  B {\bf 76}, 477 (1974).
  %%CITATION = NUPHA,B76,477;%%
  

\bibitem{Kuzenko2010}
S.~M.~Kuzenko,
``Lectures on nonlinear sigma-models in projective superspace,''
J. Phys. A {\bf 43}, 443001 (2010)
%doi:10.1088/1751-8113/43/44/443001
[arXiv:1004.0880 [hep-th]].

 \bibitem{WB} J.~Wess and J.~Bagger,
{\it Supersymmetry and Supergravity},
Princeton University Press, Princeton, 1992.


\bibitem{BMST} 
I.~Bandos, L.~Martucci, D.~Sorokin and M.~Tonin,
``Brane induced supersymmetry breaking and de Sitter supergravity,''
JHEP {\bf 1602}, 080 (2016)
[arXiv:1511.03024 [hep-th]].

\bibitem{IvanovK}
E.~A.~Ivanov and A.~A.~Kapustnikov,
``Geometry of spontaneously broken local $N=1$ supersymmetry in superspace,''
Nucl. Phys. B \textbf{333}, 439 (1990).
%doi:10.1016/0550-3213(90)90046-G

%\cite{Das:1977pu}
\bibitem{DZ2} 
  S.~Deser and B.~Zumino,
  ``Broken supersymmetry and supergravity,''
  Phys.\ Rev.\ Lett.\  {\bf 38}, 1433 (1977).
  %%CITATION = PRLTA,38,1433;%%

\bibitem{LR}
U.~Lindstr\"om and M.~Ro\v{c}ek,
``Constrained local superfields,''
Phys.\ Rev.\  D {\bf 19}, 2300 (1979).

  
 \bibitem{BFKVP} 
  E.~A.~Bergshoeff, D.~Z.~Freedman, R.~Kallosh and A.~Van Proeyen,
  ``Pure de Sitter supergravity,''
  Phys.\ Rev.\ D {\bf 92}, no. 8, 085040 (2015)
  Erratum: [Phys.\ Rev.\ D {\bf 93}, no. 6, 069901 (2016)]
  [arXiv:1507.08264 [hep-th]]. 


\bibitem{BK} I.~L.~Buchbinder and S.~M.~Kuzenko,
{\it Ideas and Methods of Supersymmetry and
Supergravity or a Walk Through Superspace},
IOP, Bristol, 1998.

\bibitem{IO}
E.~A.~Ivanov and V.~I.~Ogievetsky,
``The inverse Higgs phenomenon in nonlinear realizations,''
Teor. Mat. Fiz. \textbf{25}, 164 (1975).

   
\bibitem{VanNieuwenhuizen:1981ae}
P.~van Nieuwenhuizen,
``Supergravity,''
Phys. Rept. \textbf{68}, 189 (1981).
   
   
\bibitem{Shifman}
M.~A.~Shifman,
{\it The Many Faces of the Superworld: Yuri Golfand Memorial Volume}, World Scientific, Singapore, 2000.
 

\bibitem{OS}
  V.~Ogievetsky and E.~Sokatchev,
  ``Structure of supergravity group,''
  Phys.\ Lett.\ B {\bf 79}, 222 (1978); 
 ``The Simplest Group of Einstein Supergravity,''
Sov. J. Nucl. Phys. \textbf{31}, 140 (1980).

\bibitem{WZ78}
J.~Wess and B.~Zumino,
 ``Superfield Lagrangian for supergravity,''
 Phys.\ Lett.\  B {\bf 74}, 51 (1978).

\bibitem{old1}
K.~S.~Stelle and P.~C.~West,
``Minimal auxiliary fields for supergravity,''
Phys.\ Lett.\  B {\bf 74},  330 (1978).

\bibitem{old2}
S.~Ferrara and P.~van Nieuwenhuizen,
``The auxiliary fields of supergravity,''
Phys.\ Lett.\  B {\bf 74}, 333 (1978).


 \bibitem{IN}
E.~A.~Ivanov and J.~Niederle,
``N=1 supergravity as a nonlinear realization,''
Phys. Rev. D \textbf{45}, 4545 (1992).
%doi:10.1103/PhysRevD.45.4545
 
\bibitem{West}
P.~C.~West,
``A geometric gravity Lagrangian,''
Phys. Lett. B \textbf{76}, 569 (1978).
%doi:10.1016/0370-2693(78)90856-0 
 
\bibitem{SW1}
K.~S.~Stelle and P.~C.~West,
``de Sitter gauge invariance and the geometry of the Einstein-Cartan theory''
J. Phys. A \textbf{12}, L205 (1979).
%doi:10.1088/0305-4470/12/8/003 

\bibitem{SW2}
K.~S.~Stelle and P.~C.~West,
``Spontaneously broken de Sitter symmetry and the gravitational holonomy group,''
Phys. Rev. D \textbf{21}, 1466 (1980).
%doi:10.1103/PhysRevD.21.1466

\bibitem{PV}
C.~R.~Preitschopf and M.~A.~Vasiliev,
``The superalgebraic approach to supergravity,'' in: {\it Theory of Elementary Particles}, 
H. Dorn, D. L\"sut and G. Weigt (Eds.), Wiley-VCH, Berlin, 1998, pp. 483--488 
[arXiv:hep-th/9805127 [hep-th]].
   
 
\bibitem{DK} 
S.~Deser and J.~H.~Kay,
``Topologically massive supergravity,''
Phys.\ Lett.\ B {\bf 120}, 97 (1983).

   
\end{thebibliography}
\end{document}